\documentclass[12pt,a4paper,dvips,reqno]{article}
\usepackage{graphicx}
\usepackage{subfigure}
\usepackage{epsfig}
\usepackage{dsfont}

\begin{document}

\def\k{ {\mathtt k}}
\def\x{ {\mathsf z}}
\def\E{ {\mathcal E}}
\def\Nc{ {\mathcal P}_c}
\def\P{ {\mathcal P}}
\def\A{ {\mathcal A}}
\def\D{ {\mathcal D}_n}
\def\B{ {\mathcal B}}
\def\C{ {\mathcal C}}
\def\S{ {\mathcal S}}
\def\M{ {\mathcal M}}
\def\N{ {\mathds N}}

\def\R{ {\mathbf R}}
\def\Real{ {\mathds R}}
\def\Sf{ {\mathbf S}}

\def\H{ {\mathsf H}}
\def\im{ {\mathtt {Im}}}
\def\re{ {\mathtt {Re}}}

\def\ndpt{{SPT}}
%%%%%%%%%%%%%%%%%%%%%%%%%%%%%%%%%%%%%%%%%%%%%%%%%%%%%%%%%%%%%%%

%%%\noindent Preprint, \today
\begin{center}

\vskip 0.5 cm

{\large \bf Can  billiard eigenstates   be approximated by  superpositions
 of plane waves?}

\vskip 1 cm

 {\bf  Boris Gutkin } 

\vskip 0.5 cm

{\small  CEA-Saclay,
Service de Physique Th\'{e}orique\\
Gif-sur-Yvette Cedex,
France}\\{\small E-mail: gutkin@spht.saclay.cea.fr }

\end{center}

\vskip 1.0 cm

\begin{abstract}

\noindent
The plane wave decomposition method (PWDM) is one of the most popular
strategies   for  numerical solution of the  quantum billiard problem.
The method is based on the assumption that  each eigenstate in a billiard 
can be approximated by a  superposition of  plane waves at a given energy.
By the classical results on the theory of differential operators  this  
can indeed be justified for  billiards in convex domains.
On the contrary, in the present work we demonstrate  that eigenstates of 
non-convex billiards, in general, cannot be approximated by any  solution 
of the Helmholtz equation regular  everywhere in  $\R^2$ (in particular,  by  
linear combinations of a finite number of plane waves  having the same energy). From this we infer that PWDM cannot be applied to  billiards in non-convex 
 domains. Furthermore, it follows from our results that unlike the properties
 of   integrable billiards, where each  eigenstate can be extended into the billiard exterior as a regular solution of the Helmholtz equation,  the 
 eigenstates of non-convex billiards,  in general, do not admit such an extension.

\end{abstract}

\vskip 1.5 cm

\section{Introduction}

\noindent The quantum billiard problem  in a domain $\Omega\subset \R^2$ is defined (in units m=1) by the Helmholtz equation 
\begin{equation}(-\Delta-\k^2)\varphi(x)=0, \qquad E=\hbar^2\k^2/2 \label{1.1}
  \end{equation} 
with  Dirichlet boundary conditions 
\begin{equation}
  \varphi(x)|_{\partial \Omega}=0. \label{1.2}
  \end{equation} 
The solutions $E_n$, $\varphi_n$ of these equations determine the energy  spectrum and the set of eigenstates of $\Omega$. Studying the  properties of 
 $(E_n, \varphi_n)$ in quantum billiards  
  has became a prototype problem in ``quantum chaos''.     
 A simple form of eqs. \ref{1.1}, \ref{1.2}    suggests a  natural way to solve them.  First, for a given  energy $E$ one  looks for a set of  solutions  $\{\psi^{(n)}(\k), \, n\in\N\}$  of  the  Helmholtz equation (\ref{1.1}) in the entire  plane (without any boundary conditions). For example,   $\{\psi^{(n)}(\k)\}$ can be chosen as   a set of   plane waves: $\{ \exp(i k_n x), \,\, |k_n|=\k,\,k_n\in\Real^2\}$,  or  as   a set of   radial waves: $\{J_n(\k r)\exp(i n \theta)\}$. Then regarding  $\{\psi^{(n)}(\k)\}$ as a basis one can  search for  solutions of  eqs. \ref{1.1}, \ref{1.2} using  the  ansatz
\begin{equation}\varphi(x)=\sum a_i \psi^{(i)}(\k, x). \label{1.3} \end{equation}
As a result,  solving  eqs. \ref{1.1}, \ref{1.2} is reduced to the algebraic problem of 
finding the coefficients $a_i$ such that the linear combination (\ref{1.3}) vanishes whenever $ x\in \partial\Omega $.

 The above approach has been widely used  both in  analytical  and numerical studies of quantum billiards.    In particular, it has been suggested by  Berry in \cite{berry1} to use the expansion (\ref{1.3}) with a Gaussian amplitude distribution  to represent  eigenfunctions of quantum systems with   fully chaotic  dynamics. This idea has been applied in  numerous works to calculate   various quantities associated with   eigenfunctions, e.g., autocorrelation functions \cite{berry1}, amplitude distributions \cite{romanbacker}, statistics of nodal domains \cite{uzy2} etc.   
The same strategy   can be also used  for a numerical solution of eqs. \ref{1.1}, \ref{1.2}. In this context it  has been first introduced by  Heller \cite{heller}  with the  application  to the  Bunimovich stadium. Since that several modification of the method have been considered in \cite{kitajzi1}, \cite{kitajzi2} and in \cite{doronheller}. Depending on the choice of the   basis    in the decomposition (\ref{1.3}) one gets, in general,  different numerical  methods for solving eqs. \ref{1.1}, \ref{1.2}. Here we will single out  the basis of  plane waves (PW), most often used in  applications. For the sake of briefness we will refer to the corresponding numerical method as plane wave decomposition method  (PWDM).

As a matter of fact, the whole strategy described above    is based on the  assumption  that the set $\{\psi^{(n)}(\k)\}$ furnishes an appropriate   basis for the expansion of    solutions of eqs. \ref{1.1}, \ref{1.2}. In other words, one can use PWDM   only if    billiard  eigenstates can be  approximated 
by linear combinations  of plane waves. That means 
 \begin{equation}||\varphi_n - \psi^{[N]}||_{L^2(\Omega)} \to 0, \mbox{ as } N \to \infty  \label{1.35}\end{equation}
for  some sequence of the states $\psi^{[N]}$ which are of the form 
 \begin{equation}\psi^{[N]} = \sum_{i=1}^N a_i e^{ik_i x},\qquad k_i\in \Real^2, \,\,\, |k_i|=\k. \label{1.45}\end{equation}
We will say that the plane wave  approximation holds for a state  $\varphi_n$ if the limit (\ref{1.35}) exists.

Up to now it has been often assumed that the PWDM can be applied to billiards of arbitrary shape.
 From the results of Malgrange \cite{malgrange} (see also \cite{hormander}) on the theory of differential operators  it is known that any solution of eq. \ref{1.1} regular in a convex open domain can be approximated  by  superpositions of plane waves with $k_i\in\mathds{C}^2$.  Moreover, since each evanescent plane wave ($\im \,k_i\neq 0$) can be approximated in a bounded domain by  plane waves with real wavenumbers  \cite{berry2}, one immediately gets: 

\

\noindent{\bf Proposition 1.} {\it Let $\Omega\subset\R^2$ be a convex bounded domain,  then any solution of eq. \ref{1.1} regular in $\Omega$ can be approximated by plane waves.} 

\

\noindent This shows that the eigenstates of a quantum billiard  $\Omega$ admit PW approximation  
inside  any convex domain $\Omega_1\subset\Omega$, see fig. 1a.  Hence, PW approximation  always  holds  for   billiard eigenstates in a local sense. Furthermore, if $\Omega$ is a convex domain one can choose $\Omega_1$ in such a way that $\partial\Omega_1$ is arbitrary close to  $\partial\Omega$. Consequently, as a simple  corollary of Proposition 1 one gets:

\

\noindent{\bf Corollary 1.} {\it Eigenstates of a convex billiard $\Omega$ can  be  approximated by  superpositions of  plane waves.}

\

\noindent  The question naturally arises whether the same property holds for eigenstates of non-convex billiards, and thus, whether the  PWDM 
 can be  actually applied to the class of non-convex billiards.   

Note that there  exists an important link 
between the PWDM and the problem of eigenstate extension in quantum billiards.  Suppose $\varphi_n$ is an   eigenstate of $\Omega$ which can be extended (as a regular solution of eq. \ref{1.1}) from $\Omega$ to a convex domain  $\Omega_{2}\supset\Omega$.  Then it follows immediately by Proposition 1 that PW approximation holds for $\varphi_n$. The example of  a billiard where each eigenstate can be continued in a convex domain is shown in fig 1.b. This is the ``cake'' billiard whose boundary consists of two concentric circle arcs connected by two segments of radii at an angle $\alpha <\pi$. In the polar coordinates $x=(r,\theta)$ the eigenstates of the  ``cake'' billiard can be written explicitly as a sum of Bessel and Neumann functions:
 \[\varphi^{(m)}_n(x)= \left( a^m_n J_{\nu_m}(\k^{(m)}_{n} r) + b^m_n Y_{\nu_m}(\k^{(m)}_n r)\right)\sin\left(\nu_m(\theta  -\theta_0)\right), \qquad \nu_m=\frac{\pi m}{\alpha}.\]
Since
the  singularity point of  $\varphi^{(m)}_n(x)$ is always at the center $O$ of the circle arcs  it is possible to extend each eigenstate into  a convex domain $\Omega_2$,  see fig 1.b. Accordingly, any eigenstate of the ``cake'' billiard can be approximated by  superpositions of PW.

On the other hand, assume that for a billiard $\Omega$ an  eigenstate  $\varphi_n$  can be expanded in a basis  $\{\psi^{(n)}\}$ (see eq. \ref{1.3}), where $\psi^{(i)}$'s are solutions of the   Helmholtz equation regular in $\R^2$ (e.g., plane waves). If furthermore, the corresponding sum (\ref{1.3}) converges everywhere in  $\R^2$ it makes sense to consider   $\varphi_n(x)$ both inside and outside $\Omega$. Such extension of $\varphi_n(x)$ into $\R^2$  
provides simultaneously  solutions for the interior Dirichlet problem (when   $x\in\Omega$) and for the exterior Dirichlet problem (when $x\in\Omega^c\equiv\R^2/\Omega$).
 Based on this observation  a connection ({\it spectral duality}) between  the interior Dirichlet and the exterior scattering problems has been suggested by Doron and Smilansky in
 \cite{uzydoron}.  
The rigorous result has been established  by Eckmann and Pillet   \cite{ep}. In most  general form ({\it weak spectral duality}) it  could be stated as follows:  $E_n$ is an eigenvalue of the interior problem if and only if 
there exists an eigenvalue $e^{-i\vartheta_n}$ of the exterior scattering matrix $S(E)$ such that  $\vartheta_n (E) \to 2\pi$ whenever $E\to E_n$. Moreover, if 
 $\vartheta_n (E_n)=2\pi$ ({\it strong spectral duality}) then the corresponding interior eigenstate $\varphi_n$ could be extended into $\R^2$  as $L^2$ functions.   Therefore if strong form of spectral duality holds for some eigenenergy $E_n$ then PW approximation holds for the corresponding eigenstate $\varphi_{n}$.
It has been explicitly shown  that strong  form  of spectral duality holds  for convex  integrable billiards \cite{dis}. However, as  has been pointed out in  \cite{ep},  strong spectral duality cannot  hold for  billiards in general.

\

\noindent {\bf Remark.} It should be pointed out  that the approximability by PW is much weaker property then  strong   spectral duality. 
As  has been explained above, strong   spectral duality implies
  PW approximation for the corresponding eigenstate. The opposite, however,  is not true:  PW approximation for an eigenstate does not imply, in general, strong spectral duality.  In fact, in \cite{berry2, ep} the examples of convex billiards (in this case  the approximation by PW is possible) have been constructed where the eigenstates extension into the exterior domain  as $L^2$ functions  is not possible.

 \section{Main results}

Let $\Omega$ be a simply connected bounded domain in $\R^2$ with a piecewise smooth boundary $\partial\Omega$. Two different billiard maps can be  associated with $\Omega$. First, the standard billiard map $\Psi$  corresponding to the motion of a pointlike particle in the interior domain. Second,  the exterior map $\Psi^c$ which corresponds to the scattering off   $\Omega$ as an obstacle, see e.g., \cite{uzy}.  
In order to define the exterior map   one can  place  $\Omega$ on a sphere $\Sf^2$ of ``infinite'' radius. Then $\Psi^c$ is a standard billiard map corresponding to the motion of a pointlike particle  in the  domain $\Sf^2/\Omega$. It should be noted that there is an essential difference between  convex and  non-convex billiards. Whenever $\Omega$ is  a convex domain 
  the  interior  map $\Psi$  determines the same dynamics as the exterior map $\Psi^c$. For any interior trajectory inside $\Omega$ there is a {\it dual} trajectory in $\Omega^c$ which travels through the same set of points on the boundary $\partial \Omega$, see fig. 2a. We will refer to  this property as {\it interior-exterior duality}. In particular, for convex billiards there is one to one correspondence between the interior and exterior periodic trajectories. For each periodic trajectory $\gamma$ its continuation $\gamma^c$  into the exterior domain will be the dual periodic trajectory of the exterior map.  On the other hand,   it is straightforward to see that in  non-convex billiards  interior-exterior duality breaks down. Generally, in a non-convex billiard $\Omega$ there exist interior periodic trajectories 
  whose extension into the exterior domain  intersects   $\Omega$ again, see fig. 2b.
Let $\gamma$ be such a trajectory and let $\gamma^c$ be its extension in the exterior. Note that $\gamma\cup\gamma^c$ is a  union  of  straight lines in $\R^2$.
Take $l\subset\gamma\cup\gamma^c$ to be a  line which intersects   the boundary $\partial\Omega$ at $2n$, $n>1$ points (for the sake of simplicity we will always assume that $n=2$). Then the intersection $\Omega\cap l$ is the union of two disconnected segments:
 $\gamma_1\subset\gamma$ and $\bar{\gamma}_1\subset\gamma^c$.
If  $\bar{\gamma}_1 $ does not belong to any periodic trajectory in $\Omega$, we will refer to $\gamma$  as {\it  single periodic trajectory} ({\ndpt}). By definition any {\ndpt} has  no dual periodic trajectory in the exterior domain. In what follows we call a non-convex billiard $\Omega$ as {\it generic} if it contains  at least one stable (elliptic) or unstable (hyperbolic) {\ndpt}. According to this terminology the ``cake'' billiard in fig. 1b is non-generic, since all its periodic trajectories are of neutral (parabolic) type.

We call a smooth function $\psi(x)$ as a {\it regular solution} of the Helmholtz equation if it solves eq. \ref{1.1} everywhere in $\R^2$. For a given energy $E$ we will denote by
$\M(E)$ the set of all regular solutions of eq. 1 and by $\M_{\mathtt{PW}}(E)\subset\M(E)$ the subset of functions which can be represented as linear combinations of  finite number of  plane waves with real wavenumbers $k_i$, $|k_i|^2=2E/\hbar^2$. 
 In particular,  $\M(E)$ includes  convergent superpositions of  plane waves (also with complex wavenumbers i.e., evanescent modes) and radial waves with the  energy $E$. 
In its crudest form the main result of the present paper can be formulated in the following way. Based on the breaking of interior-exterior duality we demonstrate that eigenstates of   a generic non-convex billiard (in general) cannot be approximated by  regular solutions of eq. \ref{1.1}. To illustrate the main idea of our approach it is instructive to consider a non-convex billiard $\Omega$  with an elliptic {\ndpt} $\gamma$. It is well known that  a sequence of quasimodes $(\tilde{\varphi}_i ,\tilde{\k}_i)$ associated with $\gamma$ can be constructed (see e.g., \cite{paul}, \cite{fed}).  Each pair  $(\tilde{\varphi}_n ,\tilde{\k}_n)$ represents an approximate   solution of eqs. \ref{1.1}, \ref{1.2} such that $\tilde{\varphi}_n$ is localized along $\gamma$. Furthermore,  in the absence of systematic degeneracies  in the  spectrum of $\Omega$ the    quasimodes $(\tilde{\varphi}_n ,\tilde{\k}_n)$ approximate (in $L^2$ sense) a sequence of real solutions $( \varphi_n ,\k_n)$ of eqs. \ref{1.1}, \ref{1.2}. For each such eigenstate ${\varphi}_n$ let us consider the corresponding Husimi function

\begin{equation} H_{\varphi_n}(\x)=|\langle \x|\varphi_n\rangle|^2, \qquad \x=(q,p): \,\,\,\,\,  q\in \Omega\label{2.1},\,\,\,\,\, |p|=\hbar\k_n ,\end{equation}
where  $\langle\x|$ denotes  a  coherent state localized at the point $\x$ of the  phase space of $\Omega$. By the definition  $H_{\varphi_n}(\x)$ is localized along $\gamma$ and exponentially small  everywhere else.
On the other hand, assume that $\varphi_n$ could be approximated  by regular solutions   of eq. \ref{1.1}. That means for any $\epsilon>0$ there is $\psi_{\epsilon}\in\M(E_n)$  such that $||\varphi_n- \psi_{\epsilon}||<{\epsilon}$, where $||\cdot||$ denotes the $L^2(\Omega)$ norm. Set $q$ be a point at $\gamma_1$ and set  $p$ be directed along $\gamma_1$. Then for $\x=(q,p)$ we have

\begin{equation} H_{\varphi_n}(\x)=\lim_{\epsilon\to 0}|\langle \x|\psi_{\epsilon}\rangle|^2 =
\lim_{\epsilon\to 0}|\langle \x|e^{-it\Delta/\hbar}\psi_{\epsilon}\rangle|^2,\label{2.2}\end{equation}
where $e^{-it\Delta/\hbar}$ is the free evolution operator in $\R^2$.
Furthermore,  in the semiclassical limit the quantum evolution of  coherent states 
is governed by the corresponding classical evolution
\begin{equation}e^{-it\Delta/\hbar}|\x\rangle=e^{it E/\hbar}|\x(t)\rangle +O(\hbar^{\infty}), \qquad \x(t)=(q(t),p). \label{2.3} \end{equation}
Plugging (\ref{2.3}) into eq. \ref{2.2} and taking  time $t$ to  be such that $q(t)=q'\in \bar{\gamma}_1$ one gets
     
\begin{equation} H_{\varphi_n}(\x) - H_{\varphi_n}(\x')=O(\hbar^{\infty}), \qquad \x'=(q',p). \label{2.4} \end{equation}
 This, however, contradicts the fact that the Husimi function   $ H_{\varphi_n}(\x)$ should be exponentially decaying  outside  $\gamma$.

The above argument can be  extended to the case of hyperbolic {\ndpt} $\gamma$ as follows. Contrary to the elliptic case  it is not possible to construct  quasimodes concentrated on  hyperbolic periodic orbits. Instead, one can use a statistical approach in that case.  By the results of Paul and  Uribe \cite{paul} it is known that the average of the Husimi functions (\ref{2.1}) 

\begin{equation} \big\langle H_{\varphi_n}(\x)\big\rangle=\frac{1}{\#\P_{c\hbar}}\sum_{E_n\in \P_{c\hbar}}|\langle \x|{\varphi}_n\rangle|^2\end{equation}
over the energy interval $\P_{c\hbar}= [E-c\hbar, E+c\hbar]$, $c>0$ depends in the semiclassical limit $\hbar\to 0$   on whether $\x$ belongs  to a periodic trajectory or not. On the other hand, as   has been explained above, if each $\varphi_n$ could be approximated by a regular solution of eq. \ref{1.1} then each $H_{\varphi_n}(\x)$ (and therefore the average $\big\langle H_{\varphi_n}(\x)\big\rangle$) would be (semiclassically) invariant along ${\gamma}_1\cup\bar{\gamma}_1$.

The preceding discussion provides an intuitive explanation why it is impossible to approximate   eigenstates of a generic non-convex billiard by a superposition of plane waves. Speaking informally our  argument says that
contrary to the real eigenstates  of non-convex billiard $\Omega$, 
any  regular solution of eq. \ref{1.1} always ``preserves''
 interior-exterior duality. In what follows we consider the $L^2(\Omega)$ norm
\begin{equation}\eta_n(\psi)= ||\varphi_n-\psi||,\end{equation}
for a solution $(\varphi_n, E_n)$ of eqs. \ref{1.1},  \ref{1.2} in $\Omega$ and an arbitrary $\psi\in \M(E_n)$. By  the definition $\eta_n(\psi)$ 
measures approximability of $\varphi_n$ by regular solutions of the Helmholtz equation. Recall that  a state $\varphi_n$ is approximable by PW if
\[\inf_{\psi\in\M_{\mathtt{PW}}(E_n) } \eta_n(\psi)=0.\]

\

\noindent{\bf Remark.} Note that  by Proposition 1 for any $\psi\in\M(E_n)$ and any $\epsilon>0$ one can always find $\psi_{\epsilon}\in\M_{\mathtt{PW}}(E_n)$ such that 
$ |\eta_n(\psi)-\eta_n(\psi_{\epsilon})|<\epsilon$. In particular this  implies

\begin{equation}\eta^{\mathsf{min}}_n\equiv\inf_{\psi\in\M(E_n) } \eta_n(\psi) =  \inf_{\psi\in\M_{\mathsf{PW}}(E_n) } \eta_n(\psi).\label{2.75}\end{equation}
In other words, an  eigenstate $\varphi_n$ can be approximated by $\psi\in\M(E_n)$ if and only if it can be approximated by PW.
Therefore,  in what follows one can always assume without lost of generality that $\psi$ belongs to  $\M_{\mathtt{PW}}(E_n)$ rather than to the set $\M(E_n)$. 

\

By Corollary 1,   $\eta^{\mathsf{min}}_n=0$ for  any eigenstate  of a convex billiard.  On the contrary, in the body of the paper we show that  for a generic non-convex billiard  the average of $\eta^{\mathsf{min}}_n$ over an  energy interval is bounded from below by a strictly positive constant:

\

\noindent{\bf Proposition 2.} {\it Let $\Omega$ be a non-convex billiard with at least one stable or unstable  {\ndpt} and let ($\varphi_n,E_n$), $n=1,2, ...\,\infty$ denote the eigenstates and eigenenergies of the corresponding quantum billiard. For any set of approximating functions $\{\psi_i\in\M(E_i), \,\, i\in \N\}$ 
the average of $\eta_n=\eta_n (\psi_n)$ over the energy interval $\P_{c\hbar}= [E-c\hbar, E+c\hbar]$,   satisfies 

\begin{equation}\big\langle\eta_n\big\rangle >  \C(\hbar),\qquad \mbox{ where }\qquad \B=\lim_{\hbar\to 0}\C(\hbar)/\hbar \label{2.7}\end{equation}
 is strictly positive and independent of $\psi_i$'s.  Moreover, if $\Omega$ contains a {\ndpt} $\gamma$ of elliptic type then (provided the spectrum of $\Omega$ has no systematic degeneracies)  
there exists an infinite subsequence $\S_{\gamma}=\{(\varphi_{j_m},E_{j_m}), \,\, m\in \N\}$ (of a positive density, i.e., $\lim_{N\to\infty}\frac{\#\{j_m|j_m<N\}}{N}>0$) 
 such that for any $(\varphi_n, E_n) \in \S_{\gamma}$ and any regular solution $\psi\in \M(E_n)$ 

\begin{equation}\eta_n(\psi) >  \C_{\gamma}+O(\hbar^{1/2}),\label{2.8}\end{equation}
where $\C_{\gamma}$ is  a strictly positive constant independent of $\psi$ and $\hbar$.}

\

\noindent 
From  (\ref{2.7},\ref{2.8}) one immediately obtains the  corollary:  

\

\noindent{\bf Corollary 2.} {\it For a generic non-convex billiard $\Omega$ there exists an infinite  subsequence 
of eigenstates $\{\varphi_{j_n}, \,\, n\in\N\}$ such that:
1) $\eta^{\mathsf{min}}_{j_n}>0$; 2) $\varphi_{j_n}$ cannot be extended into the domain $\Omega^c$ (as  a regular solution of eq. \ref{1.1}).}

\

\noindent Obviously, this implies the following properties of a generic non-convex billiard:

\begin{itemize}
\item In general,    eigenstates of non-convex billiards do not admit  approximation by PW and PWDM cannot be used in that case;

\item  The spectral duality for a generic non-convex billiard holds only in the weak form. 
\end{itemize}

The paper is organized as follows. In the next section we collect several necessary facts  about   coherent states.
In Sec. 4  the case of   elliptic  {\ndpt}'s   is considered. First, using the coherent states
we construct a family of quasimodes  $(\tilde{\varphi}_n,\tilde{E}_n)$ associated with such  trajectories. Then, we show that the lower bound  (\ref{2.8}) holds for the  eigenstates $\varphi_n$ approximated by $\tilde{\varphi}_n$.
The case of  hyperbolic  {\ndpt}'s is considered in Sec. 5. Here we use the results of  Paul and  Uribe  to  estimate  the average $\big\langle \eta_n\big\rangle $  over an energy interval. Finally in Sec. 6 we discuss our results and consider possible generalizations.

%%%%%%%%%%%%%%%%%%%%%%%%%%%%%%%%%%%%%%%%%%%%%%%%%%%%%%%%%%%%%%%%%%%%%%%%%%%%%%%

 \section{Coherent  states}

\noindent{\bf Definition of  coherent states.}  The coherent states  have been introduced already in the beginning of quantum mechanics and have been used in many areas since then. The basic idea is to built a complete set of vectors of Hilbert space  localized in the phase space  both in $q$ and $p$ directions at the scale $\sqrt{\hbar}$.  The standard example of such states in $\R^d$ is given by the  Gaussians:

\begin{equation} u_{\x}^{\sigma}(x)=\left({\det \im \sigma}\right)^{\frac{1}{4}}\left({\hbar\pi}\right)^{-\frac{d}{4}}e^{\frac{i}{\hbar}[\langle p,x-q\rangle+\frac{1}{2}\langle x-q, \sigma\, (x-q)\rangle]}, \,\,\,\, \x=(q,p), \,\,\,\, \im \sigma>0. \label{3.1} \end{equation}
In the present work we will consider a slightly more general class of coherent states. (For a more
general definition of coherent states see e.g., \cite{paul}.) Let $\rho^{\varepsilon}_q(\cdot)$ be a  $C_{0}^{\infty}$ function in $\R^d$ equal to one in a neighborhood of the point $q$ and zero outside  the  sphere of radius $\varepsilon$  centered at $q$. A coherent state at $\x=(q,p)$   is the vector 

\begin{equation}\phi^{\sigma}_{\x} (x)=\rho^{\varepsilon}_q(x) u_{\x}^{\sigma}(x).\label{3.2}\end{equation} 
It is easy to see that  the  coherent states (\ref{3.2}) are semiclassicaly orthogonal:
\begin{equation} ||\phi^{\sigma}_{\x}||^2=1+O(\hbar), \,\,\, \langle\phi^{\sigma}_{\x}|\phi^{\sigma}_{\x'}\rangle=O(\hbar^{\infty})\mbox{ if }\x\neq\x'.\label{3.4}\end{equation}
The role of the cut-off $\rho^{\varepsilon}_q(x)$ is rather technical, it allows to define   coherent states  inside  compact domains.  To use the vectors (\ref{3.2}) as coherent states inside a billiard domain $\Omega$ one needs that 
\begin{equation}\mbox{supp}[\rho^{\varepsilon}_q(x)]\subset\Omega. \label{3.3}\end{equation}

\

\noindent{\bf Propagation of  coherent  states.}
 An important property of  coherent states is that their quantum evolution in the semiclassical limit is completely determined by the corresponding classical evolution. Let $\H=-\hbar^2\Delta/2+v(x)$ be the operator of symbol $\mathcal{H}=p^2/2 + v(x)$  inducing the  flow $\Psi^t:V\to V$ on the phase space $V$. Then, as it is well known, for any time $t$ the propagation of the coherent state $\phi^{\sigma}_{\x}$ localized at $\x\in V$ is given by
\begin{equation}   e^{-i t \H/\hbar}\phi^{\sigma}_{\x} = e^{i\left({ S(t)/\hbar}+\mu(t)\right)}  \phi^{\sigma(t)}_{\x(t)} +O(\hbar^{1/2}), \label{3.5}\end{equation}
where $S(t)=\int_0^t \left(p \dot{q}-\mathcal{H}(p,q)\right) dt$ is the classical action along the path $\x(t)$ and  $\mu(t)$ is the Maslov index.
The parameters $ \x(t)=\Psi^t\cdot\x$, $\sigma{(t)}=D\Psi^t \cdot\sigma$ in eq. \ref{3.5} are determined by the  evolution of the  initial data  $ \x$,  $\sigma$ under the flow $\Psi^t:\x\to\x(t)$ and its derivative   
 \begin{equation}D\Psi^t :\sigma\to\sigma(t)=\frac{a\sigma+b}{c\sigma+d} ,\end{equation}
where $d\times d$ matrices $a,b,c,d$ are the components of $D\Psi^t$ in a given coordinate system: 
\[D\Psi^t =\left(\begin{array}{cc}
a & b \\
c & d \\ \end{array}\right).\] It is convenient to chose two of $2d$ coordinates in
the phase space $V$ to be along the flow and along the line orthogonal to
the energy surface. Then the matrix $\sigma$ can be decomposed into
$\sigma=\sigma^{0}\oplus\sigma^{1}$, where  the scalar part $\sigma^{0}$ corresponds to the above two directions and $(d-1)\times(d-1)$ matrix $\sigma^{1}$ corresponds to the orthogonal subspace. It is straightforward to see that in such a basis  $ D\Psi^t$ acts separately on  $\sigma^{1}$ and   $\sigma^{0}$. In particular,   $D\Psi^t \cdot\sigma^0$ is given by a linear transformation:

\begin{equation}D\Psi^t \cdot\sigma^0=\frac{\sigma^0}{u\sigma^0+1}.\end{equation}

In the present paper we will use the above results for two types of two-dimensional flows:  free evolution on $\R^2$  under the Hamiltonian $\H_0$ ($v(x)=0$)  and the evolution induced by the billiard  Hamiltonian $\H_{\Omega}$ ($v(x)=0$ if $x\in\Omega$ and $v(x)=\infty$ otherwise).  
Let us consider in  some detail the evolution of coherent states in  billiards. 
Set $\Omega$ be the  billiard domain. We will denote by $\Psi_{\Omega}^t: V\to V$ the billiard flow, whose action is on the  standard phase space $V$ of $\Omega$. It should be pointed out  that  one can use the coherent states (\ref{3.2}) for the point $\x=(q,p)\in V$ only if $q$ is sufficiently far away from the boundary  $\partial\Omega$. Indeed, to satisfy the condition (\ref{3.3})  $q$ has to  be  at the distance larger than $\varepsilon$ from the boundary. 
For the sake of simplicity, we will not consider a generalized class of coherent states defined in the whole  domain $\Omega$, rather we will use the states (\ref{3.2}) but only for the interior points of
$\Omega$.
For this purpose let us define the  inner domain $\Omega_{\varepsilon}\subset\Omega$ which contains all the points $q$ of $\Omega$ such that  the distance between $q$ and $\partial\Omega$ is larger then $\varepsilon$: $\mbox{dist}(q,\partial\Omega)\geq \varepsilon$, see fig. 3.
In what follows we  will  fix $\varepsilon$ to be a  small compare to  linear sizes of the billiard (but large compare to $\hbar^{1/2}$) and consider the coherent states propagation   under the condition  that   at the initial moment $t_1=0$ and the  final moment $t_2=t$  the  points  $\x(0)$, $\x(t)$ belong to the domain  $\Omega_{\varepsilon}$. Whenever this condition is fulfilled one can use the formula (\ref{3.5}), where the states $\phi^{\sigma}_{\x}, \phi^{\sigma(t)}_{\x(t)}$ are both of the form (\ref{3.2}). Furthermore, if $q(t)\in\Omega_{\varepsilon}$ for all $t\in [t_1,t_2]$ (i.e., there is  no collisions with the boundary between the times $t_1$ and $t_2$)
then    the reminder term in (\ref{3.5}) is of the order $O(\hbar^{\infty})$.

\

\noindent{\bf Husimi functions.} Let $\varphi_n$ be an eigenstate of $\H$ with the  eigenenergy $E_n$. Given a  coherent state $\phi^{\sigma}_{\x}$ one can construct the corresponding Husimi function:

\begin{equation}  H_{n}(\x)=|\langle\phi^{\sigma}_{\x}|\varphi_n\rangle|^2\qquad \x=(q,p); \,\,\, \sigma=(\sigma^0,\sigma^1), \,\, -i\sigma^0=\beta >0.\label{3.6}\end{equation} 
Based on the propagation formula (\ref{3.5}) the following average over Husimi functions \begin{equation}
 \sum_n f (\omega_n)| \langle  \varphi_n|\phi^{\sigma}_{\x} \rangle|^2
=\sum_{l=0}^{\infty} d_l \, \hbar^{\frac{1}{2}+l}, \,\,\, \omega_n=\frac{E_n -E}{\hbar}, \,\,\, E=p^2/2 \label{3.7}\end{equation}
 has been calculated to the leading order by  Paul and  Uribe \cite{paul}. It turns out that the result depends on whether the classical trajectory through $\x$ is periodic or not. With the  application to the Hamiltonian $\H_{\Omega}$ the results in \cite{paul} read as follows. Let $\tilde{f}(\cdot)$ be the Fourier transform of $f$.
If $\x$ is not periodic under the   flow $\Psi_{\Omega}^t$ then 

\begin{equation} d_{0}=\left(\frac{1}{\beta E}\right)^{1/2} \tilde{f}(0).\label{3.8} \end{equation}
Alternatively, if $\x$ belongs to a  periodic trajectory  the  additional terms (of the  same order in $\hbar$) arise.  In particular, for a hyperbolic periodic trajectory $\gamma$ with the period $T_{\gamma}$ the leading term in (\ref{3.7}) is given by
\begin{equation} d_{0}= \left(\frac{1}{\beta E}\right)^{1/2}\left( \sum_{l=-\infty}^{+\infty} \tilde{f}(l T_{\gamma})
\frac{e^{il( S_{\gamma}/\hbar +\mu_{\gamma} )} }{\cosh^{1/2}(l\lambda_{\gamma})} \right
) , \label{3.9}\end{equation}
where 
\[S_{\gamma}=2ET_{\gamma}, \,\, \,\, \mu_{\gamma}, \,\,  \,\, \lambda_{\gamma} \]
are the action, Maslov index and Laypunov exponent of $\gamma$.

%%%%%%%%%%%%%%%%%%%%%%%%%%%%%%%%%%%%%%%%%%%%%%%%%%%%%%%%%%%%%%%%%%%%%

\section{PW approximation  for eigenstates of non-convex billiards (elliptic case)}

\noindent Let $\gamma$ be a periodic orbit in the billiard $\Omega$ and let $\Gamma(E)$ be the ``lift'' of $\gamma$ to the phase space $V$ at the energy $E$. This means $\Gamma(E)$ is a set of the points $\x=(q,p)\in V$ such that $q\in\gamma$, $p^2=2E$ and the  vector $p$ is directed along $\gamma$. Obviously, for any $\x \in\Gamma(E)$,  \, $\Psi^{ T_{\gamma}}_{\Omega}\cdot\x =\x$, where $T_{\gamma}$ is the period of the trajectory. We will make use of the letter $\varepsilon$ to denote the restriction of $\gamma$, $\Gamma(E)$  to the domain $\Omega_{\varepsilon}$ i.e., $\gamma^{\varepsilon}=\{q\in\gamma\cap\Omega_{\varepsilon}\}$, $\Gamma^{\varepsilon}(E)=\{\x=(q,p)\in\Gamma(E): \,\, q\in\Omega_{\varepsilon}\}$. Provided that  $\gamma$ is elliptic a set of approximate solutions ({\it quasimodes}) $ \tilde{\varphi}_n(x)$  of eqs. \ref{1.1}, \ref{1.2} associated with $\gamma$  can be constructed. The possibility of  quasimode construction  on elliptic periodic orbits is well known. In the following we will  follow the approach developed in \cite{paul'}, \cite{paul} (see also \cite{roman}, \cite{paul''} and the references there). 

Before we turn to the construction of the states $ \tilde{\varphi}_n(x)$ in billiards let us  recall a general definition for  quasimodes.

\

\noindent {\bf Definition.} Let $H$ be a Hilbert space  and $\H$ be  a self adjoint operator  with the domain $D(H)$. A pair $(\tilde{\varphi}_n,\widetilde{E}_n)$ with $\tilde{\varphi}_n\in D(H)$, $||\tilde{\varphi}_n||=1$ and $\widetilde{E}_n\in \R$ is called a quasimode with the discrepancy $\delta_n$, if 

\begin{equation} (\H - \widetilde{E}_n)\tilde{\varphi}_n=r_n, \,\,\,\mbox{ with }\,\,\, ||r_n||=\delta_n.\label{4.0.1}\end{equation}

\

\noindent By a general theory  (see e.g., \cite{laz}) the  quasimodes  $(\tilde{\varphi}_n,\widetilde{E}_n)$ should be close to an exact solution $(\varphi_n,E_n)$  of the equation 
 \begin{equation} (\H - {E})\varphi=0 \label{4.0.2} \end{equation}
in the following  sense. If  $(\tilde{\varphi},\widetilde{E})$ is a quasimode with the discrepancy $\delta$ then  there exists at least one eigenvalue of $\H$ in the interval 
   \begin{equation}\P_{\delta}=[\widetilde{E}-\delta, \widetilde{E}+\delta].\label{4.0.3}\end{equation}
Furthermore, let $\nu$ be the  distance between $\widetilde{E}$ and an eigenvalue $E_i$ of $\H$ outside $\P_{\delta}$, then
   \begin{equation}||\tilde{\varphi}-\pi_{\nu}\, \tilde{\varphi}||\leq \frac{\delta}{\nu},\label{4.0.4}\end{equation}
where $\pi_{\nu}$ denotes the spectral projection operator on the part of the spectrum  $\{E_n\}$ inside the interval $(\widetilde{E}-\nu, \widetilde{E}+\nu)$.

\

\noindent{\bf Remark.} In general, the formula (\ref{4.0.4}) implies that any state $\tilde{\varphi}_n$ approximates a  superposition of eigenstates $\varphi_n$. In order to approximate  individual eigenstates of $\H$,    $\delta_n$ should be much less than the  energy intervals: $\Delta E_{n}=|E_n -E_{n+1}|$, $\Delta E_{n-1} =|E_n -E_{n-1}|$. 
For  two dimensional billiards  $\big\langle\Delta E_{n}\big\rangle \sim  \hbar^2 $, so the approximation  of ${\varphi}_n$ by $\tilde{\varphi}_n$ becomes semiclassically ($\hbar\to 0$) meaningful only if the spectrum of $\Omega$ has no systematic degeneracies and  quasimodes with discrepancy $\delta \sim \hbar^{\alpha}$, $\alpha > 2$ can be constructed. For the quantum billiard problem a quasimode  construction providing   $\delta=O(\hbar^{\infty})$ is known to exist \cite{popov} and for the rest of this section we will   assume that  the billiard spectrum has no systematic degeneracies.

\subsection{Quasimode construction}

We will now schematically describe the construction of quasimodes 
concentrated on elliptic periodic
orbits. The basic idea is to lunch a coherent state along the orbit and average over the time.  As it can be shown, this procedure yields an approximately invariant state  if the initial state is chosen in the right way, see e.g.,  \cite{paul,roman}. Let $\phi^{\sigma}_{\x }$, $\x=(q, p)\in\Gamma^{\varepsilon}(E)$ be a coherent state localized on the periodic orbit $\gamma$. We will associate with $\gamma$  the state

\begin{equation} |\Phi^{\sigma}_{\Gamma(E)}\rangle= \frac{1}{C}\int_{0}^{T_{\gamma}} e^{i t (E-\H_{\Omega}) /\hbar}|\phi^{\sigma}_{\x }\rangle \,dt, \label{4.1.1}\end{equation} where $C$ is fixed by the normalization condition 
$||\Phi^{\sigma}_{\Gamma(E)}||=1$
and $T_{\gamma}$ is the  period of  the classical evolution along $\gamma$:
$\x(T_{\gamma})=\x$. 
 The propagation formula (\ref{3.5})  yields 

\begin{eqnarray}(E-\H_{\Omega})\Phi^{\sigma}_{\Gamma(E)}&=&{r_{\gamma}}, \nonumber \\
C r_{\gamma}
&=&i\hbar\left(e^{i(S_{\gamma}/\hbar+\mu_{\gamma})}\phi^{\sigma(T_{\gamma})}_{\x }-\phi^{\sigma}_{\x }\right) +O(\hbar^{3/2}),
\label{4.1.2}\end{eqnarray}
where $S_{\gamma}$, $\mu_{\gamma}$ are the classical action and Maslov index after one period.
Therefore, $C r_{\gamma}=O(\hbar^{3/2})$ provided that the following conditions are satisfied:

\

\noindent {\bf Condition  1:}  $\sigma(T_{\gamma})=\sigma$;  \,\, {\bf Condition  2:} $S_{\gamma}/\hbar+\mu_{\gamma}= 2\pi n$ for some integer $n$. 

\

\noindent  For each $n$ let $\E_n$, $\sigma_n=(\sigma^0_n,\sigma^1_n)$ denote solutions of Conditions 1, 2. It  is possible to show (see e.g., \cite{paul}) that the first condition can be satisfied if and only if $\sigma^0_n=0$ and $\gamma$ is an elliptic periodic orbit. The second condition impose the Bohr-Sommerfeld quantization on the quasienergy $\E_n$. When both conditions are satisfied the corresponding pair $(\E_n,\Phi^{\sigma_n}_{\Gamma(\E_n)} )$ provides the  quasimode with the discrepancy $\delta_{\gamma}=O(\hbar^{3/2})/C$.

\

\noindent{\bf Remark.} It should be noted that a much wider class of quasimodes concentrated on $\gamma$ can be constructed by this method if one uses in (\ref{4.1.1})  coherent states with transverse  excitations \cite{paul,paul''}. For  simplicity of exposition, we restrict our consideration only  to the  quasimodes without transverse excitations, whose leading order is determined by eq. \ref{4.1.1}.

\

 To construct quasimodes with discrepancies of higher order in $\hbar$ one has to consider the time evolution of  coherent states of a more general type. This leads to  transport equations whose solvability  pose additional conditions  on the quasienergies, see \cite{roman}. From the results of Cardoso and Popov \cite{popov} the prossibillity to construct  quasimodes $(\widetilde{E}_n,\tilde{\varphi}_n)$ in billiards having  discrepancy  $\delta_{\gamma}= O(\hbar^{\infty})$ is known to exist.
Let $(s, y)$ be  a coordinate system in a neighborhood of $\gamma$ such that   $s$ is a coordinate along the trajectory and $y$ is a coordinate in the orthogonal direction.
Using these coordinates
 the leading order of   $(\widetilde{E}_n,\tilde{\varphi}_n)$  can be written as  follows \cite{fed, roman}:
\begin{equation} \widetilde{E}_n={\E}_{n}+O(\hbar^2), \qquad 
 \tilde{\varphi}_n(x)=e^{iv(x)/\hbar}u(x)+O(\hbar),\label{4.1.4} \end{equation}
where 
\[ 
v(s,y)=v_0(s)y^2+O(y^3), \qquad
u(s,y)=u_0(s)+O(y^2)  \]
and the  parameters $v_0(s)$, $u_0(s)$ are
   determined    by    Conditions 1, 2:
\begin{equation}\Phi^{\sigma_n}_{\Gamma(\E_n)}(x)=e^{iv_0(s)y^2/\hbar}u_0(s),\qquad x=(s, y).\label{4.1.45} \end{equation}

 As has been explained before, in the absence of systematic degeneracies in  the billiard spectrum one can expect that,  in general, a state  $\tilde{\varphi}_n$ approximates an individual eigenstate of the billiard $\Omega$. In what follows we will denote by  $\widetilde{\S}_{\gamma}$  the set of quasimodes for which  $\tilde{\varphi}_n$    approximates some eigenstate $ \varphi_n$  (rather than  a linear combination of $ \varphi_n$'s) and by ${\S}_{\gamma}$ the set of true solutions of eqs. \ref {1.1}, \ref{1.2} corresponding to $\widetilde{\S}_{\gamma}$.  Then by eq. \ref{4.0.4} for each  $(\tilde{\varphi}_i,\widetilde{E}_i)\in\widetilde{\S}_{\gamma} $ and $({\varphi}_i,E_i)\in{\S}_{\gamma} $ we have   
 
\begin{equation} \C^1_i=||\tilde{\varphi}_i-\varphi_i||=O(\hbar^{\infty}), \qquad |\widetilde{E}_i -E_i|=O(\hbar^{\infty}) .\label{4.1.5} \end{equation}

\subsection{A lower bound for the approximation of eigenstates}

The quasimode construction described in the previous section is quit general and can be applied to an arbitrary elliptic periodic trajectory. 
In the present section we will consider eigenstates of the billiard $\Omega$ from the subset ${\S}_{\gamma}$, where $\gamma$ is an elliptic  {\ndpt}.
We show that for $(\varphi_n, E_n)\in {\S}_{\gamma}$ and any regular solution $\psi\in\M(E_n)$ of eq. \ref{1.1} in $\R^2$ the norm

   \begin{equation}  \eta_n(\psi)=||\varphi_n -\psi||  \label{4.2.1} \end{equation}
is bounded from below by  
\begin{equation}
\eta_n(\psi) \geq {\C}_{\gamma} +   {\mathcal C}^1_n +
O(\hbar^{1/2}),\label{4.2.2}\end{equation}
 where ${\mathcal C}_{\gamma}$ is a positive constant determined only by  geometrical parameters   of the periodic orbit. Since ${\C}^1_n = O(\hbar^{\infty})$, this implies  the inequality ($\ref{2.8}$) holds for any $(\varphi_n, E_n)\in {\S}_{\gamma}$.

Let $\gamma$ be an elliptic {\ndpt} and let $\gamma_1$,  $\bar{\gamma}_1$ be as  defined in Sec. 2, see fig. 3. Now  fix the parameter $\varepsilon$ to be sufficiently small such that $\gamma^{\varepsilon}_1\equiv\gamma_1\cap\Omega_{\varepsilon}\neq\emptyset$,  $\bar{\gamma}^{\varepsilon}_1\equiv\bar{\gamma}_1\cap\Omega_{\varepsilon}\neq\emptyset$. We will denote by the capital letters   $\Gamma_1(E)$,  $\bar{\Gamma}_1(E)$ (resp. $\Gamma^{\varepsilon}_1(E)$,  $\bar{\Gamma}^{\varepsilon}_1(E)$)   the corresponding ``lifts'' of  $\gamma_1$,  $\bar{\gamma}_1$ (resp. $\gamma^{\varepsilon}_1$,  $\bar{\gamma}^{\varepsilon}_1$)  into the phase space $V$ at the energy shell $E$.  Recall that the main idea  behind the quasimode construction  (\ref{4.1.1})  is to use  coherent states propagating along a periodic orbit.
By analogy, one can construct  states localized on $\gamma_1$ and $\bar{\gamma}_1$. 
Let  $\x(0)=\x \in \Gamma_1(E)$. Consider the classical evolution (both for positive and negative time) of $\x$ under the free flow $\Psi^t_0:\x\to\x(t)=(q(t),p(t))$ in $\R^2$. Obviously, as  time evolves, the point $q(t)$ successively crosses the boundary of $\Omega_{\varepsilon}$ at the sequence of points $q_1, q_2,\bar{q}_1,\bar{q}_2$,  see fig. 3. We will denote by  $t_1, t_2, \bar{t}_1,\bar{t}_2$  the corresponding time moments:   $q_1=q(t_1), q_2=q(t_2),\bar{q}_1=q(\bar{t}_1),\bar{q}_2=q(\bar{t}_2)$.  Then the states localized along  $\gamma_1$ and $\bar{\gamma}_1$ are given by

\begin{equation} |\Phi^{\sigma}_{\Gamma_1(E)}\rangle= \int^{t_1}_{t_2} e^{it (E-\H_0) /\hbar}|\phi^{\sigma}_{\x }\rangle \,dt, \label{4.2.3}\end{equation} 
\begin{equation} |\Phi^{\sigma}_{\bar{\Gamma}_1(E)}\rangle= \int^{\bar{t}_1}_{\bar{t}_2} e^{it (E-\H_0) /\hbar}|\phi^{\sigma}_{\x }\rangle \,dt. \label{4.2.4}\end{equation} 
Note, that under the  free evolution $e^{-it \H_0 /\hbar}$ the support of $\phi^{\sigma}_{\x }$ is not preserved inside  $\Omega$, and therefore
   the supports of  $\Phi^{\sigma}_{\Gamma_1}, \Phi^{\sigma}_{\bar{\Gamma}_1}$ do not belong to the billiard domain. However, one can  slightly modify the definition of the states   $\Phi^{\sigma}_{\Gamma_1}, \Phi^{\sigma}_{\bar{\Gamma}_1}$  to make them admissible as billiard states in  $\Omega$. Let $\x=\x_1$, $\sigma=\sigma_1$ be as before and set $\tau$ be  such  that under the classical evolution $\Psi^{\tau}_0:\x_1\to\x(\tau)$ the point $\x(\tau)=\x_2$ belongs to ${\bar{\Gamma}^{\varepsilon}_1}$. Set $\phi^{\sigma_2}_{\x_2 }(x)= e^{-i\tau \H_0 /\hbar}\phi^{\sigma}_{\x }(x)+O(\hbar^{\infty})$ be  the coherent state in $\Omega$, whose parameters are given by: $(\sigma_2, \x_2)= (D\Psi^{\tau}_0 \cdot\sigma_1, \Psi^{\tau}_0\cdot\x_1)$.
 Then the states 
\begin{eqnarray} |\bar{\Phi}^{\sigma}_{\Gamma_1(E)}\rangle&=& \int^{t_1}_{t_2} e^{it (E-\H_{\Omega}) /\hbar}|\phi^{\sigma_1}_{\x_1 }\rangle \,dt, \label{4.2.5}\\ 
|\bar{\Phi}^{\sigma}_{\bar{\Gamma}_1(E)}\rangle
&=& \int^{\bar{t}_1-\tau}_{\bar{t}_2-\tau} e^{it (E-\H_{\Omega}) /\hbar}|\phi^{\sigma_2}_{\x_2 }\rangle \,dt, \label{4.2.6}\end{eqnarray}
have their supports in $\Omega$ and satisfy 
\begin{equation} |\bar{\Phi}^{\sigma}_{\Gamma_1(E)}\rangle= 
|\Phi^{\sigma}_{\Gamma_1(E)}\rangle
+O(\hbar^{\infty}), \qquad  
|\bar{\Phi}^{\sigma}_{\bar{\Gamma}_1(E)}\rangle
 =|\Phi^{\sigma}_{\bar{\Gamma}_1(E)}\rangle
+O(\hbar^{\infty}).\label{4.2.7}\end{equation}

 To  get the lower bound (\ref{4.2.2}) we are going first to  construct   a  state ${\Phi}$  with the  property 

\begin{equation} \langle \psi  |{\Phi}\rangle=0+ O(\hbar^\infty),\label{4.2.9} \end{equation}
for any $\psi\in\M(E')$. Let us show how such a state can be constructed using  $\bar{\Phi}^{\sigma}_{\Gamma_1}$, $\bar{\Phi}^{\sigma}_{\bar{\Gamma}_1}$. Set $\langle\cdot|\cdot\rangle_{\R^2}$,  $\langle\cdot|\cdot\rangle$ be the scalar products  in $L^2(\R^2)$ and $L^2(\Omega)$ respectively.  
 From the definitions (\ref{4.2.3},\ref{4.2.4}) one has

\begin{equation}\langle \psi |\Phi^{\sigma}_{\Gamma_1(E)}\rangle_{\R^2}
=\int^{t_1}_{t_2} e^{it (E-E') /\hbar}\langle \psi |\phi^{\sigma}_{\x }\rangle \,dt
={C_1(\omega)}\langle \psi |\phi^{\sigma}_{\x }\rangle, \label{4.2.10} \end{equation} 
where
\begin{equation} C_{1}(\omega) = \exp{\left( \frac{i(t_1+t_2) \omega}{2} \right) }\, \frac{2 \sin(\omega T_{\gamma_1}/2)} { \omega }, \,\, T_{\gamma_1}=|t_1-t_2|, \label{4.2.11}\end{equation} 
and $ \omega = {(E - E')}/{\hbar}$. Analogously: 
 
\begin{equation}
\langle \psi |\Phi^{\sigma}_{\bar{\Gamma}_1(E)}\rangle_{\R^2}={C_2(\omega)}\langle \psi |\phi^{\sigma}_{\x }\rangle,\label{4.2.12}\end{equation}
with 

\begin{equation} C_{2}(\omega) = \exp{\left( \frac{i(\bar{t}_1+\bar{t}_2) \omega}{2} \right) }\, \frac{2 \sin(\omega T_{\bar{\gamma}_1}/2)} { \omega }, \,\, T_{\bar{\gamma}_1}=|\bar{t}_1-\bar{t}_2|    .\label{4.2.13} \end{equation} 
Furthermore, let us introduce the states

\begin{equation} |\Phi^{\sigma}_{1}(E,E')\rangle=\frac{1}{C_1(\omega)}|\bar{\Phi}^{\sigma}_{\Gamma_1(E)}\rangle, \qquad |\Phi^{\sigma}_{2}(E,E')\rangle=\frac{1}{C_2(\omega)}|\bar{\Phi}^{\sigma}_{\bar{\Gamma}_1(E)}\rangle.\label{4.2.135} \end{equation} 
 Then it follows immediately from eqs. \ref{4.2.10}, \ref{4.2.12} that the state $\Phi ={\Phi}^{\sigma}(E,E')$,

\begin{equation}|{\Phi}^{\sigma}(E,E')\rangle=|\Phi^{\sigma}_{1}(E,E')\rangle -|\Phi^{\sigma}_{2}(E,E')\rangle
 \label{4.2.14} \end{equation} 
satisfies orthogonality condition (\ref{4.2.9}).
 
Let $({\varphi}_n, E_n)\in\S_{\gamma}$ be a solution of eqs. \ref{1.1}, \ref{1.2} 
and let $(\tilde{\varphi}_n,\tilde{E}_n)\in\widetilde{\S}_{\gamma}$ be the corresponding quasimode, whose leading order parameters $\E_n$, $\sigma_n=(\sigma^0_n,\sigma^1_n)$  are determined by Conditions 1, 2, see eqs. \ref{4.1.4}, \ref{4.1.45}. 
Now fix the energy parameters in eq. \ref{4.2.14} by   $E=\E_n$, $E'=E_n$ and put $\sigma=\bar{\sigma}_n$, where $\bar{\sigma}_n=(i\beta, \sigma^1_n)$ and $\beta$ is an arbitrary real positive  number.
We will make use of the state 
\[|{\Phi_n}\rangle=|\Phi^{\bar{\sigma}_n}(\E_n,E_n)\rangle\]
 in order  to get a lower bound on $\eta_n$.  
For any $\psi\in \M(E_n)$  we have   

 \begin{equation}
||\tilde{\varphi}_n-\psi|| \, ||{\Phi}_n || \geq
|\langle\tilde{\varphi}_n-\psi|{\Phi}_n\rangle| =
|\langle\tilde{\varphi}_n|{\Phi}_n\rangle|+O(\hbar^{\infty}).
\label{4.2.15}\end{equation}
 Using  the triangle inequality 

 \begin{equation} ||\tilde{\varphi}_n-\varphi_n || + ||\varphi_n-\psi ||\geq  ||\tilde{\varphi}_n-\varphi_n +\varphi_n-\psi || =||\tilde{\varphi}_n-\psi||\label{4.2.16} \end{equation}
 one  gets immediately  from (\ref{4.2.15})

\begin{equation} \eta_n(\psi)=||\varphi_n-\psi|| \geq   \frac{ |\langle\tilde{\varphi}_n|\Phi_n \rangle|}{  || \Phi_n || } - {\mathcal C}^1_n+O(\hbar^{\infty}). \label{4.2.17} \end{equation}
It remains to estimate the scalar product $|\langle\tilde{\varphi}_n|\Phi_n \rangle|$ and the norm of the state ${\Phi_n}$. First, consider the norm $||{\Phi_n}||$. Since $ {\gamma}_1\cap{\bar{\gamma}_1}=\emptyset$ one has from the definition of $\Phi_n$

\begin{equation}\langle\Phi_n |\Phi_n\rangle=
\frac{1}{|C_1(\omega_n)|^2}\langle{\Phi}^{\bar{\sigma}_n}_{\Gamma_1(\E_n)} |{\Phi}^{\bar{\sigma}_n}_{\Gamma_1(\E_n)}\rangle+\frac{1}{|C_1(\omega_n)|^2}
\langle{\Phi}^{\bar{\sigma}_n}_{\bar{\Gamma}_1(\E_n)}|
{\Phi}^{\bar{\sigma}_n}_{\bar{\Gamma}_1(\E_n)}\rangle+O(\hbar^{\infty}),\label{4.2.18}  \end{equation}
with $\omega_n=(E_n -\E_n)/ \hbar$.
The calculations of the scalar products performed in Appendix give: 
\begin{eqnarray} \langle{\Phi}^{\bar{\sigma}_n}_{\Gamma_1(\E_n)} |{\Phi}^{\bar{\sigma}_n}_{\Gamma_1(\E_n)}\rangle&=& T_{\gamma_1} \left(\frac{2\pi\hbar}{\beta E_n}\right)^{1/2} +O(\hbar) , \nonumber \\
\langle{\Phi}^{\bar{\sigma}_n}_{\bar{\Gamma}_1(\E_n)}|
{\Phi}^{\bar{\sigma}_n}_{\bar{\Gamma}_1(\E_n)}\rangle
 &=& T_{\bar{\gamma}_1}  \left(\frac{2\pi\hbar}{\beta E_n}\right)^{1/2}+O(\hbar)\label{4.2.19} \end{eqnarray} 
and  for the leading order of  $C_1(\omega_n)$, $C_2(\omega_n)$  one has from  
 eqs. \ref{4.2.11}, \ref{4.2.13}
\begin{equation}|C_2(\omega_n)| = T_{\bar{\gamma}_1} +O(\hbar),\qquad |C_1(\omega_n)|= T_{\gamma_1}  +O(\hbar).\label{4.2.20} \end{equation}
Combining (\ref{4.2.19}) and (\ref{4.2.20}) together  one finally  gets
\begin{equation}\langle \Phi_n |\Phi_n\rangle=  \left(\frac{2\pi\hbar}{\beta E_n}\right)^{1/2}
\left( 
\frac{1}{ T_{\gamma_1}}
 + \frac{1}{ T_{\bar{\gamma}_1}}
\right) +O(\hbar). \label{4.2.21} 
\end{equation} 
In the same way for the scalar product $\langle\tilde{\varphi}_{n} |\Phi_{n}\rangle$ we have by (\ref{4.1.4},\ref{4.1.45})

\begin{eqnarray}|\langle\tilde{\varphi}_{n} |\Phi_n\rangle|=
 | \langle\Phi^{\sigma_n}_{\Gamma(\E_n)} |\Phi^{\bar{\sigma}_n}_{\Gamma_1(\E_n)}\rangle|+O(\hbar) 
&=& \frac{T_{\gamma_1}}{ T_{\gamma} }|\langle\Phi^{\sigma_n}_{\Gamma(\E_n)} |\Phi^{\sigma_n}_{\Gamma(\E_n)}\rangle|^{1/2}  | \langle\Phi^{\bar{\sigma}_n}_{\Gamma_1(\E_n)}|\Phi^{\bar{\sigma}_n}_{\Gamma_1(\E_n)}
\rangle|^{1/2} \nonumber\\  
+  O(\hbar)  
&=& \frac{1}{T_{\gamma}}\left(\frac{2\pi\hbar}{\beta E_n}\right)^{1/2}+O(\hbar).
\label{4.2.22} \end{eqnarray}
The estimation (\ref{4.2.2}) follows now immediately after inserting eqs. \ref{4.2.21}, \ref{4.2.22} into (\ref{4.2.17}). The resulting constant  $\C_{\gamma}$, which determines the lower bound on $\eta_n$  in the semiclassical limit reads as
 \begin{equation}
 {\C}_{\gamma}=\sqrt{ \frac{ T_{\bar{\gamma}_1} T_{\gamma_1} }
{(T_{\bar{\gamma}_1}+ T_{\gamma_1}) T_{\gamma} } }=
\sqrt{ \frac{ \ell_{\bar{\gamma}_1} \ell_{\gamma_1} }
{(\ell_{\bar{\gamma}_1}+ \ell_{\gamma_1}) \ell_{\gamma} } } +O(\varepsilon),\label{4.2.23} 
 \end{equation}
where  $\ell_{\bar{\gamma}_1}, \ell_{\gamma_1}, \ell_{\gamma} $ are the lengths of ${\bar{\gamma}_1}, {\gamma_1}$ and ${\gamma} $ respectively.

%%%%%%%%%%%%%%%%%%%%%%%%%%%%%%%%%%%%%%%%%%%%%%%%%%%%%%%%%%%%%%%%%%

\section{PW approximation  for eigenstates of non-convex billiards (hyperbolic case)}

\noindent In the present section we consider the case of   a hyperbolic  {\ndpt} $\gamma$. Let as before $\{\varphi_n(x)\}$ be the set of eigenfunctions in ${\Omega}$ approximated by regular solutions $\{\psi_n(x)\}$ of eq. \ref{1.1}. For an arbitrary set of $\psi_n(x)\in\M(E_n)$, $n=1,2,...\,\infty$ we will estimate the average of 
\begin{equation}\eta_n\equiv\eta_n(\psi_n)=||\varphi_n -\psi_n||\label{5.1}\end{equation}
over an energy interval. Our objective is to show that independently of the choice
of $\psi_n$'s, in  the limit $\hbar\to 0$
 the average $\big\langle\eta_n\big\rangle$  is bounded from below by a strictly positive  constant.

Let $\Phi^{\sigma}_{1}(E,E')$, $\Phi^{\sigma}_{2}(E,E')$, $\Phi^{\sigma}(E,E')$ be as in the previous section with the parameter $\sigma$ of the form $\sigma=(i\beta, \sigma^1)$, $\beta>0$. 
 For each integer $n$ we will consider the states
\begin{equation}|\Phi_{n,1}\rangle=|\Phi^{\sigma}_{1}(E,E_n)\rangle, \qquad
|\Phi_{n,2}\rangle=|\Phi^{\sigma}_{2}(E,E_n)\rangle \label{5.2}\end{equation}
and their difference
\begin{equation}|\widetilde{\Phi}_{n}\rangle=|\Phi_{n,1}\rangle-|\Phi_{n,2}\rangle=|\Phi^{\sigma}(E,E_n)\rangle, \label{5.25}\end{equation}
 which is   orthogonal to any $\psi\in\M(E_n)$ up to the term $O(\hbar^{\infty})$  (see eq. \ref{4.2.9}).
In addition, it will be  also useful to introduce  the state 
\begin{equation}|\widetilde{\Phi}'_{n}\rangle=|\Phi_{n,1}\rangle+|\Phi_{n,2}\rangle.\label{5.3}\end{equation}
Note that $\widetilde{\Phi}'_{n}$ is orthogonal to $\widetilde{\Phi}_{n}$  in the  semiclassical limit.

 Similarly to the  case of elliptic {\ndpt}'s, one can make use of the state $\widetilde{\Phi}_n$ to get a lower bound on $\eta_n$: 

\begin{equation} \eta_n \geq \frac{|\langle\widetilde{\Phi}_n|
 \,\varphi_n-\psi_n \rangle|} {||\widetilde{\Phi}_n||}= \frac{|\langle\widetilde{\Phi}_n|\varphi_n \rangle|} {||\widetilde{\Phi}_n||}+O(\hbar^{\infty}).\label{5.4} 
 \end{equation}
In order to estimate the right side of this inequality let us consider the
 following difference 

\begin{equation}\D=
 |\langle{\Phi}_{n,1}|
\varphi_n \rangle|^2 - |\langle{\Phi}_{n,2}|
\varphi_n \rangle |^2.  \label{5.5}
\end{equation}
Using the states  $\widetilde{\Phi}_n$, $ \widetilde{\Phi}'_n$
one can rewrite  $\D$ as

\begin{equation}\D=\re\left(  \langle\widetilde{\Phi}_n|\varphi_n \rangle \langle\widetilde{\Phi}'_n|\varphi_n \rangle^{*}\right).\label{5.6}\end{equation}
Hence,
the following inequality follows immediately 

\begin{equation}
|\D|  \leq  |\langle\widetilde{\Phi}_n|\varphi_n \rangle| \, | \langle\widetilde{\Phi}'_n|\varphi_n \rangle| \leq ||\widetilde{\Phi}'_n||\,\, |\langle\widetilde{\Phi}_n|\varphi_n \rangle|. \label{5.7}
\end{equation}
Finally, since $||\widetilde{\Phi}_n||-||\widetilde{\Phi}'_n||=O(\hbar^{\infty})$, we get by (\ref{5.4}) and (\ref{5.7})

\begin{equation} \eta_n \geq \frac{|\D|}{||\widetilde{\Phi}_n||\,||\widetilde{\Phi}'_n||}+O(\hbar^{\infty}) = \left| \frac{ |\langle{\Phi}_{n,1}|
\varphi_n \rangle|^2 - |\langle{\Phi}_{n,2}|
\varphi_n \rangle |^2 }{\langle\widetilde{\Phi}_n|\widetilde{\Phi}_n\rangle}\right|+O(\hbar^{\infty}) .\label{5.8}
\end{equation}

We will now use this inequality  to  get a lower bound for the sum of $ \eta_n $
over the energy interval $\P_{c\hbar}=[E-c\hbar, E+c\hbar]$, where $c$ is a positive constant. One has straightforwardly  from (\ref{5.8})

\begin{equation}\sum_{E_n \in\P_{c\hbar} } \eta_n > \left| \sum_{E_n \in \P_{c\hbar}} \frac{ |\langle{\Phi}_{n,1}|
\varphi_n \rangle|^2 }{\langle\widetilde{\Phi}_n|\widetilde{\Phi}_n\rangle}
-  \sum_{E_n \in\P_{c\hbar}  } \frac{ |\langle{\Phi}_{n,2}|
\varphi_n \rangle|^2 }{\langle\widetilde{\Phi}_n|\widetilde{\Phi}_n\rangle}
\right|+O(\hbar^{\infty}).\label{5.9}\end{equation}
Furthermore,  the definition of the states ${\Phi}_{n,1}$,  ${\Phi}_{n,2}$ implies
\begin{equation}
|\langle{\Phi}_{n,1}|\varphi_n \rangle|^2= |\langle\phi^{\sigma_1}_{\x_1}|\varphi_n \rangle|^2, \,\, \x_1\in\Gamma^{\varepsilon}_1; \,\,\,\, 
|\langle{\Phi}_{n,2}|\varphi_n \rangle|^2= |\langle\phi^{\sigma_2}_{\x_2}|\varphi_n \rangle|^2, \,\, \x_2\in\bar{\Gamma}^{\varepsilon}_1, \label{5.10}
\end{equation}
where  $(\x_1, \sigma_1)=(\x, \sigma)$ and $(\x_2, \sigma_2)=(\x(\tau), \sigma(\tau))$ are  related by the free classical evolution as in the previous section.
 As a result, the inequality (\ref{5.9}) reads as

\begin{equation}\sum_{E_n \in\P_{c\hbar} } \eta_n >
\left|\sum_n f (\omega_n)| \langle  \varphi_n|\phi^{\sigma_1}_{\x_1} \rangle|^2
-\sum_n f (\omega_n)| \langle  \varphi_n|\phi^{\sigma_2}_{\x_2} \rangle|^2\right|+O(\hbar^\infty), \label{5.11}
\end{equation}
with $ \omega_n={(E-E_n)}/{\hbar}$ and 
\[
 f(\omega_n)=\left\{\begin{array}{cl}
{1/\langle\widetilde{\Phi}_n|\widetilde{\Phi}_n\rangle} & \mbox{if $\omega_n \in [-c,c]$} \\
0 & \mbox{otherwise}.
\end{array} \right.\]
The elementary  calculations (see Appendix)  provide the leading order of the function $f(\omega_n)$, $\omega_n \in [-c,c]$:
\begin{eqnarray}
 f(\omega_n)&=&\frac{1}{{\langle{\Phi_{n,1}}|{\Phi_{n,1}}\rangle} + {\langle{\Phi_{n,2}}|{\Phi_{n,2}}\rangle}  }+O(\hbar^{\infty})\nonumber \\ 
&=&\frac{2|p|}{(\pi\hbar\beta)^{\frac{1}{2}}  } \left( \frac{\omega^2_n T_{\gamma_1}}{\sin^2(\omega_n T_{\gamma_1}/2)}+
\frac{\omega^2_n T_{\bar{\gamma}_1}}{\sin^2(\omega_n T_{\bar{\gamma}_1}/2)}
\right)^{-1}+O(\hbar^{0}). 
\label{5.12}
\end{eqnarray}

 Now  we can apply to
(\ref{5.11}) the results of  Paul and  Uribe (see Sec. 3). Taking into account  that  $\x_1\in\Gamma$ while $\x_2$ does not belong to any periodic trajectory,  we get by eqs. \ref{3.8}, \ref{3.9} 
 the following estimation for the average of $\eta_n$:
  
\begin{equation}\big\langle\eta_n\big\rangle\equiv\frac{1}{\#\P_{c\hbar} }\sum_{E_n \in \P_{c\hbar} } \eta_n >
 \frac{1}{\#\P_{c\hbar} } \left| \sum_{l\neq 0} \widetilde{F}(l T_{\gamma})
\frac{e^{il( S_{\gamma}/\hbar +\mu_{\gamma} )} }{\cosh^{1/2}(l\lambda_{\gamma})} \right| +O(\hbar^{3/2}), \label{5.13} \end{equation}
where $ \widetilde{F}(\cdot)$ is the Fourier transform of the function

\[F(x)=\left\{\begin{array}{cl} \left(\frac{8}{\pi}\right)^{\frac{1}{2}}
\left( \frac{x^2 T_{\gamma_1}}{\sin^2(x T_{\gamma_1}/2)}+
\frac{x^2 T_{\bar{\gamma}_1}}{\sin^2(x T_{\bar{\gamma}_1}/2)}
\right)^{-1} & \mbox{if $x \in [-c,c]$} \\
0 & \mbox{otherwise}
\end{array} \right.\]
and $\#\P_{c\hbar}$ is the number of eigenstates in the interval $\P_{c\hbar}$
whose leading order  for a billiard of  area ${\mathcal A}$ is
 given by the  Weyl formula:  \[\#\P_{c\hbar}={\mathcal A} c/2\pi\hbar+O(\hbar^0).\]   Consequently, if 
\begin{equation} Y=\left| \sum_{l\neq 0} \widetilde{F}(l T_{\gamma})
\frac{e^{il( S_{\gamma}/\hbar +\mu_{\gamma} )} }{\cosh^{1/2}(l\lambda_{\gamma})} \right| \neq 0  \label{5.14}\end{equation}
one has from (\ref{5.13})  
\begin{equation}\big\langle\eta_n\big\rangle \,\, > \B \hbar +O(\hbar^{3/2}), \,\,\,\, \,\, \,\, \B=2\pi Y/c{\mathcal A}>0.  \label{5.15}\end{equation} 
If moreover one assumes that $T_{\bar{\gamma}_1}c, \, T_{{\gamma}_1} c<<1$, the function $F(x)$ takes a simple form:

\[F(x)\approx\left\{\begin{array}{cl} \left(\frac{1}{2\pi}\right)^{\frac{1}{2}}
\left( \frac{T_{\bar{\gamma}_1}T_{\gamma_1} }{T_{\bar{\gamma}_1}+ T_{\gamma_1} } \right) & \mbox{if $x \in [-c,c]$} \\
0 & \mbox{otherwise}
\end{array} \right.\]
and the constant $\B$ can be written  explicitly:

\begin{equation}\B\approx\frac{ \sqrt{2\pi} }{\mathcal A} 
\left( \frac{T_{\bar{\gamma}_1}T_{\gamma_1} }{T_{\bar{\gamma}_1}+ T_{\gamma_1} } \right) 
\left| \sum_{l\neq 0} \frac{\sin(l c T_{\gamma})}{l c T_{\gamma}} \,\,
\frac{e^{il( S_{\gamma}/\hbar +\mu_{\gamma} )} }{\cosh^{1/2}(l\lambda_{\gamma})}
 \right| . \label{5.16}\end{equation}

Note, that the lower bound (\ref{5.15}) has been obtained using only one {\ndpt}. In the case of  hyperbolic dynamics, however, the periodic orbits (and, in particular,  {\ndpt}'s) proliferate exponentially. Therefore, one can improve the estimation  (\ref{5.15}) making use of a state $\widetilde{\Phi}_n^{{\sf sum}}$  which is concentrated on a set of   {\ndpt}'s $\{\gamma\}$ and satisfies eq. \ref{4.2.9}. A simple way to construct such a state is to define it as the superposition:
  \begin{equation}\widetilde{\Phi}_n^{{\sf sum}}=\sum_{\{\gamma\}}\widetilde{\Phi}_n({\gamma}),\label{5.17}\end{equation}
where $\widetilde{\Phi}_n({\gamma})$ stands for the  state  (\ref{5.25}) associated with a {\ndpt} $\gamma$.

 Finally, let us  mention   that the statistical estimation (\ref{5.15}) can be straightforwardly generalized to the case of elliptic {\ndpt}'s. In that case  one should use the analogs of eqs. \ref{3.8}, \ref{3.9}   (which are known to exist \cite{paul}) for stable periodic trajectories.

%%%%%%%%%%%%%%%%%%%%%%%%%%%%%

\section{Discussion and conclusions}

Speaking informally, Proposition 2 implies that there is no on-shell  basis  of regular solutions of the Helmholtz equation which can be used to approximate   all eigenstates  of a generic non-convex billiard.  That means
any linear combination of plane waves, radial waves etc., with the same energy fails to approximate real eigenstates of non-convex billiards.
 In fact, a  stronger result can be shown.  Let $\Omega$ be  a  generic non-convex billiard and let 
$\Omega'$ be a domain (not necessarily convex) which  properly contains $\Omega$: $\Omega'\supset\Omega$,  $\partial\Omega'\cap\partial\Omega=\emptyset$. Denote by $\M_{\Omega'}(E)$ the set of all solutions of eq. \ref{1.1} regular in $\Omega'$ (note, that $\M_{\Omega'}(E)\supseteq\M(E)$). Let us argue that the eigenstates of $\Omega$ cannot be approximated, in general, by states belonging to $\M_{\Omega'}(E)$. Let  $\gamma$ be  a {\ndpt} and let $l,\gamma_1,\bar{\gamma}_1$  be as defined  before. Furthermore, assume that the  segment of the line $l$ between  $\gamma_1$ and $\bar{\gamma}_1$  is entirely in  $\Omega'$, see fig. 4. (It seems to be a natural assumption that in a generic case one can always fined such a  {\ndpt},   provided $\Omega'$ properly contains $\Omega$). 
 Then take $\Omega_0\subset\Omega$ to be a convex domain satisfying:  $\Omega_0\cap\gamma_1\neq\emptyset$, $\Omega_0\cap\bar{\gamma}_1\neq\emptyset$. Now, suppose  an eigenstate $\varphi_n$ of $\Omega$ can be
approximated by  states $\psi'(x)$ from $\M_{\Omega'}(E_n)$.
According to  Proposition 1 $\psi'(x)$ can be approximated in  ${\Omega_0}$ by  regular solutions of eq. \ref{1.1} and thus for any ${\epsilon}>0$  there exists $\psi_{\epsilon}\in\M(E_n)$ such that  $||\varphi_n(x)-\psi_{\epsilon}(x)||_{L^2(\Omega_0)}<{\epsilon}$. Therefore, applying the same arguments as in Sec. 2 we get 
\[H_{\varphi_n}(\x_{1})-H_{\varphi_n}(\x_{2}) =\lim_{\epsilon\to 0}|\langle\x_{1}|\psi_{\epsilon}\rangle|^2 -\lim_{\epsilon\to 0}|\langle\x_{2}|\psi_{\epsilon}\rangle|^2=O(\hbar^{\infty}),\]
where $\x_{1}=(q_1,p)\in\Gamma_1(E_n)$, $q_1\in\gamma_1\cap\Omega_0$
and  $\x_{2}=(q_2,p)\in\bar{\Gamma}_1(E_n)$,  $q_2\in\bar{\gamma}_1\cap\Omega_0$. However, as has been pointed out before, this cannot be true for each $n$  since $\x_{2}\notin\Gamma$.

 The  two   properties of generic non-convex billiards follows  immediately from the above analysis.  First, it is not possible to approximate eigenstates of a generic non-convex billiard $\Omega$ also if one includes in the basis $\{\psi^{(n)}(\k)\}$  singular solutions of eq. \ref{1.1}, e.g., the Hankel functions
\[ \{ H_{n}^{\pm}(k|x-x_i|)e^{in\theta(x,x_i)}, \, n\in \N \}, \]
with a finite number of singularity points $x_i$. Second, there exists an infinite sequence of eigenstates which do not admit extension into  any large  domain $\Omega'$ properly  containing $\Omega$.
 That means the continuation of the interior eigenstates of  a generic non-convex billiard into the exterior domain should be (in general) impossible because of singularities which occur arbitrary close to the billiard's boundary. It remains as an open problem what is  the exact nature  of such singularities. (For example, whether one can, in principal, extend  eigenstates beyond the  boundary of a generic non-convex
billiard.) It  should be also mentioned that the problem of  the eigenstates extension in convex billiards is beyond the scope of the present paper. It would become a natural question to  inquire about the relation between the billiard  shape  and the  type of singularities arising for the extended eigenstates. In particular, it would be interesting to know whether  the strong form of spectral duality (when it is possible to  extend eigenfunctions in $\R^2$ as regular solutions of the Helmholtz equation) holds exclusively for   integrable billiards.

 Further, let us stress an important difference between the cases of elliptic and hyperbolic dynamics. The counting function $\mathcal{N}^*(\k)=\#\{\tilde{\k}_n<\k\}$ for quasimodes $(\tilde{\varphi}_n,\tilde{\k}_n )$ which can be constructed on an elliptic periodic trajectory  is known to be of the same asymptotic form $\mathcal{N}^*(\k)=\alpha\k^{2}+ O(\k)$, $\alpha>0$ as the  counting function $\mathcal{N}={\mathcal A}\k^{2}/4\pi+ O(\k)$ for the real spectrum $\{\k_n\}$,  see \cite{popov}.  
 Therefore, in a generic case, if an elliptic {\ndpt} $\gamma$ exists the subsequence $\{\varphi_{j_n}, n\in \N\}$ of billiard eigenstates approximated by the quasimodes concentrated on  $\gamma$ should be   of  the positive density: 
\[\lim_{N\to\infty}\frac{1}{N}\#\{j_n|j_n\leq N\}=\lim_{\k\to\infty}\frac{\mathcal{N}^*(\k)}{\mathcal{N}(\k)}>0.\]
Since for each $ \varphi_{j_n}$ the estimation (\ref{2.8}) holds, that means there   exists a  subsequence    of  eigenstates with a positive density which do not admit  approximation  by plane waves.  
 In the case of hyperbolic dynamics the statistical lower bound  (\ref{2.7})  implies,  in fact, only a weaker result. It says that an infinite sequence (possibly of zero density) of such states exists. However, if one assumes that all eigenstates of fully chaotic billiards have ``uniform properties'' the inequality (\ref{2.7}) suggests a natural conjecture:

\

\noindent{\bf Conjecture.} {\it For a  non-convex billiard with fully chaotic
dynamics the set of states which can be approximated by PW is of density zero.}

\

\noindent Note, that it is impossible  to exclude the possibility of  existence of ``exceptional''    eigenstates (the eigenstates which can be approximated by PW) in non-convex billiards. Indeed, one can take a finite superposition of plane waves $\psi^{[N]}$ and  set  a (non-convex) nodal domain of  $\psi^{[N]}$ to be the billiard's boundary. Then $\psi^{[N]}$ itself is the eigenstate of this billiard which can be approximated by PW.

Finally, the study of the present paper is restricted to the two-dimensional simply connected  domains with Dirichlet boundary conditions. However, it is easy to see that  presented results allow several rather straightforward generalizations. First, higher dimensional billiards and different types of boundary conditions can be treated in the same way. Second, billiards in multiply connected domains (fig. 5)  have the same properties as non-convex billiards. Consequently, all the results obtained for non-convex billiards hold for multiply connected billiards as well. Third, we conjecture that our results can be generalized to the billiards on non-compact manifolds with non-trivial metrics (also in the presence of a potential) e.g., billiards on the hyperbolic plane. In such a case, one needs to  adjust the notion of domain's ``convexity'' to the corresponding classical dynamics. In other words, a domain should be defined as ``convex'' if the interior-exterior duality holds and defined as ``non-convex'' if it breaks dawn. 

%%%%%%%%%%%%%%%%%%%%%%%%%%%%%%%%%%%%%%%%%%%%%%%%%%%%%%%%%%%%%%%%%%%%%%%%%%%%%%%

\noindent \section*{Acknowledgments}

I am grateful to  R. Schubert, U. Smilansky, A. Voros and S. Nonnenmacher
for useful discussions.     

%%%%%%%%%%%%%%%%%%%%%%%%%%%%%%%%%%%%%%%%%%%%%%%%%%%%%%%%%%%%%%%%%%%%%%%%%%%%%%%

\section*{Appendix}

{\bf Proposition 3.} {\it Let  $\Phi^{\sigma}_{\Gamma}$, $\Phi^{\bar{\sigma}}_{\Gamma}$ be the states:
 \begin{eqnarray} |\Phi^{\sigma}_{\Gamma}\rangle &=&  \frac{1}{C_1}\int_{0}^{T}e^{i(E-\H_0)t/\hbar} |\phi^{\sigma}_{\x}\rangle dt, \qquad \sigma=(\sigma^0,\sigma^1),  \nonumber \\ 
|\Phi^{\bar{\sigma}}_{\Gamma}\rangle &=&  \frac{1}{C_2}\int_{0}^{T}e^{i(E-\H_0)t/\hbar} |\phi^{ \bar{\sigma}}_{\x}\rangle dt, \qquad \bar{\sigma}=( \bar{\sigma}^0, \bar{\sigma}^1)
 \label{A.1}\end{eqnarray}
 localized along the path $\Gamma=\Gamma(E)$, $\Gamma(E)=\{ \Psi^t \cdot \x=(q(t),p(t)),\, t\in [0, T] , E=p^2/2\}$ with   $\sigma^0=i\beta_1$,  $\bar{\sigma}^0=i\beta_2$; $\beta_1,\beta_2 >0$ and  $\sigma^1=\bar{\sigma}^1$. Then
 \begin{equation} \langle \Phi^{\sigma}_{\Gamma} |\Phi^{\sigma}_{\Gamma}\rangle =\frac{T}{C_1^2}\left(\frac{2\pi\hbar}{\beta_1 E}\right)^{1/2} +O(\hbar); \,\,\,\,
\langle \Phi^{\bar{\sigma}}_{\Gamma} |\Phi^{\bar{\sigma}}_{\Gamma}\rangle =\frac{T}{C_2^2}\left(\frac{2\pi\hbar}{\beta_2 E}\right)^{1/2} +O(\hbar),
 \label{A.2} \end{equation}
 \begin{equation} \langle \Phi^{\sigma}_{\Gamma} |\Phi^{\bar{\sigma}}_{\Gamma}\rangle = \langle \Phi^{\sigma}_{\Gamma} |\Phi^{\sigma}_{\Gamma}\rangle^{1/2} \langle \Phi^{\bar{\sigma}}_{\Gamma} |\Phi^{\bar{\sigma}}_{\Gamma}\rangle^{1/2} +O(\hbar). \label{A.25} \end{equation} 
}

\

\noindent{\it Proof.} The inner product

\begin{equation}\langle \Phi^{\sigma}_{\Gamma} |\Phi^{\bar{\sigma}}_{\Gamma}\rangle =\frac{1}{ C_1 C_2}\int_{0}^{T} \int_{0}^{T} \langle\phi^{\sigma}_{\x}|   e^{i(E-H_0)(t_1-t_2)/\hbar} |\phi^{\bar{\sigma}}_{\x} \rangle dt_{1} dt_{2} \label{A.3}\end{equation}
can be written as 

\begin{eqnarray}  \langle \Phi^{\sigma}_{\Gamma} |\Phi^{\bar{\sigma}}_{\Gamma}\rangle&=& \frac{1}{2 C_1 C_2 }\left(\int_{0}^{T}(T-t)H(t) dt 
 + \int_{0}^{T}(T-t)H(-t) dt \right) \nonumber \\ 
&=&\frac{1}{2 C_1 C_2  }\int_{-T}^{T}(T-|t|)H(t) dt, \label{A.4} \end{eqnarray}
where 
\begin{equation} H(t)=\langle\phi^{\sigma}_{\x}|   e^{i(E-H_0)t/\hbar} |\phi^{\bar{\sigma}}_{\x} \rangle.\label{A.5}\end{equation}
By the propogation formula (\ref{3.5}) we get for  (\ref{A.5})

\begin{eqnarray} H(t)&=& e^{i(S(t)+E)/\hbar +i\mu(t)}\langle\phi^{\sigma}_{\x}|\phi^{\bar{\sigma}(t)}_{\x(t)} \rangle+O(\hbar)\nonumber \\ 
&=&\det\left( \frac{4\im \sigma\im \bar{\sigma}^*(t)}{(\sigma-\bar{\sigma}^*(t))^2}\right)^{1/4}\exp\left(-\frac{i t^2}{2\hbar} \langle p , \bar{\sigma}^*(t)\frac{1}{\sigma-\bar{\sigma}^*(t)}\sigma \, p  \rangle \right) +O(\hbar) \nonumber \\
&=&\left(\frac{(\beta_2 \beta_1)^{1/4}}{(\beta_2+ \beta_1)^{1/2}}+O(t)\right)\exp\left(-\frac{ t^2 p^2\beta_2 \beta_1}{2\hbar(\beta_2 +\beta_1)} 
 +O(t^3)\right)+O(\hbar).
\end{eqnarray}
After inserting this expresion into eq. \ref{A.4} and applying the stationary phase approximation to the integral one  gets (\ref{A.2},\ref{A.25}). Finaly, let us note  that eq. \ref{A.25} remains true also when $\beta_1$ or $\beta_2$ equals zero.

\end {document}
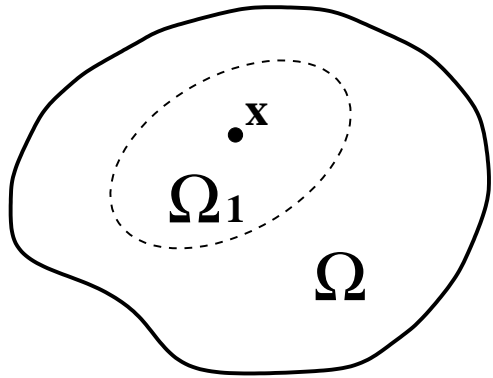
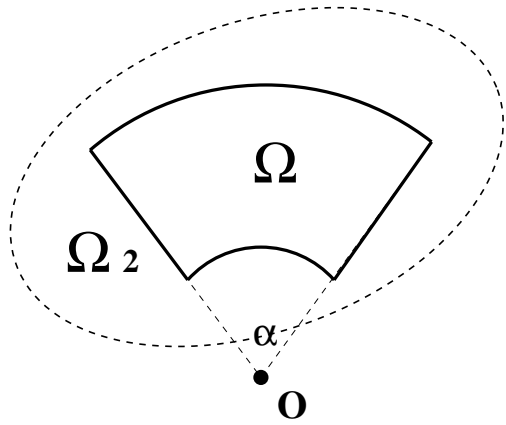
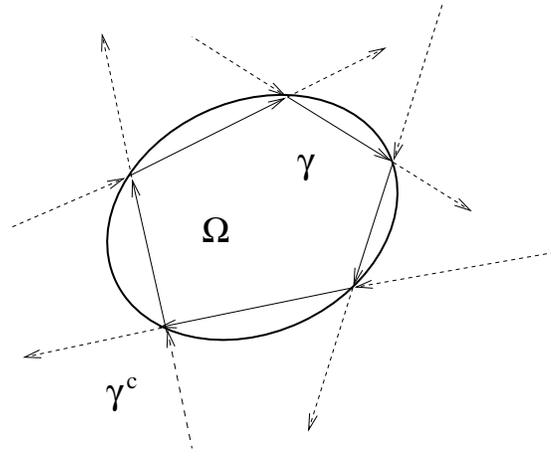
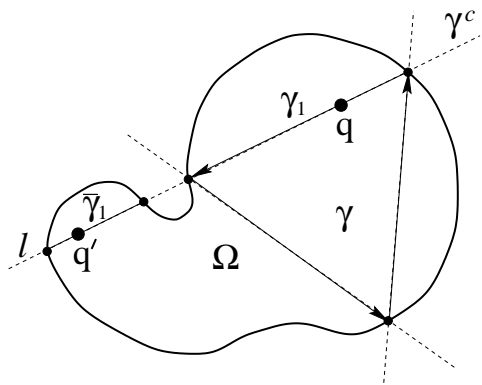
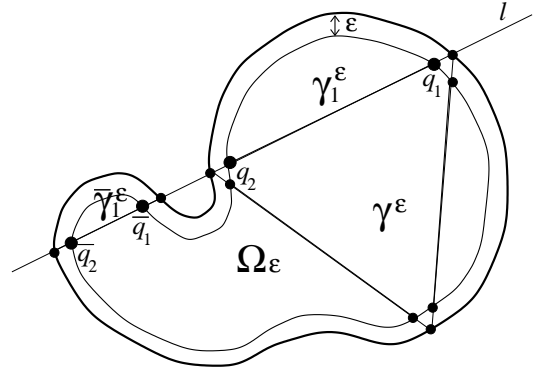
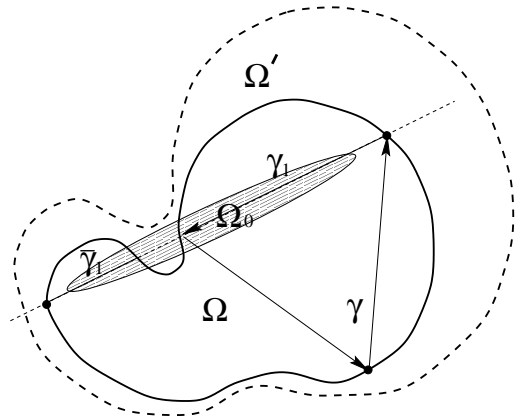
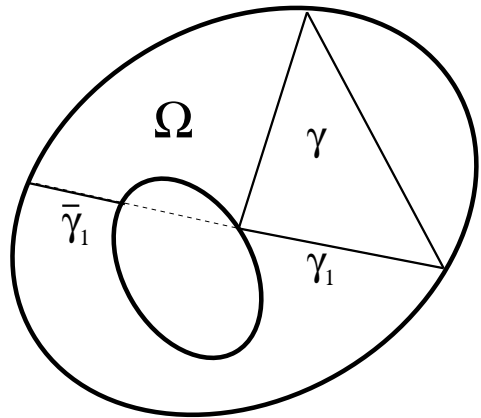
